\documentclass[11pt]{article}
\usepackage{natbib}
\usepackage{latexsym}

\begin{document}

\bibliographystyle{apalike.bst}

\baselineskip 24pt

%
%  The following commands are for short-cuts
%

\newcommand{\ket}[1]{|#1\rangle}
\newcommand{\bra}[1]{\langle#1|}
\newcommand{\opa}[1]{\mbox{\boldmath $#1$}^{\ast}}
\newcommand{\ip}[2]{\langle#1|#2\rangle}
\newcommand{\braket}[2]{\langle#1|#2\rangle}
\newcommand{\op}[1]{\mbox{\boldmath $#1$}}
\newcommand{\lat}{\mathcal{L}}
\newcommand{\alg}[1]{\mathcal{#1}}
\newcommand{\tpv}[2]{\ket{#1}\otimes \ket{#2}}
\newcommand{\tp}[2]{#1\otimes #2}
\newcommand{\ketbra}[2]{|#1\rangle\langle#2|}

\newenvironment{proof}{{\it Proof.}}{\mbox{$\Box$}}
\newtheorem{theorem}{Theorem}
\newtheorem{definition}{Defintiion}

\pagenumbering{arabic}

\title{Discussion:\\Are There Material Objects in Bohm's Theory?}

\author{Michael Dickson}

\maketitle

\section{The Issue}

\citet{bedard99a} argues that ``Bohm's interpretation is not as
classical as it initially appears'' (p.~223) and that, in particular,
the common view that the pilot-wave
theory\footnote{Louis \citet{debroglie24a} formulated essentially the same
theory that \citet{bohm52a} did.  An examination of his
dissertation reveals that in fact his theory was much closer to Bohm's
than is usually assumed.  I have chosen, therefore, to use the
`neutral' term `pilot-wave theory'.} can get by with an ontology of
{\it just} particles ``do[es] not make sense'' (ibid.).  Indeed,
``sets of Bohmian particles do not have all the intrinsic properties
necessary to constitute a material object'' (ibid.).  This last remark
is the heart of Bedard's impressive paper, the point being that the
`minimalist' interpretation of the pilot-wave theory (according to
which the only entities are the particles, and their only fundamental
property position) has no account of material objects.

In this note, I shall suggest that in fact the minimalist's
interpretation is fully adequate.  Bedard shows that the minimalist's
physical world (consisting of just particles with position) is too
sparse to explain why, for example, a collection of particles is ``a
brain instead of something merely shaped [as] a brain'' (ibid.).
Instead, the wavefunction itself must be invoked to get any account of
the forces that bind particles together.  The heart of my response is
just that the minimalist is not obligated to have such an account.
Much of the rest of this note is aimed at making this claim
plausible. I hasten to add that it is no part of my argument that the
pilot-wave theory is `classical', whatever that adjective may mean.

\section{Review of the Arguments}

Bedard presents three arguments against the minimalist interpretation
of the pilot-wave theory, according to which the only `real' objects
are particles, and the only `real' properties are positions, while the
wave function does nothing more than encode the dynamics obeyed by
those particles.  I shall review each argument, reserving most
comments for later.

Her first argument is based on the idea that ``in order for a system to
constitute a composite object such as a cat, table, or hammer, the
right types of particles must be bonded together in an appropriate
way'' (p.~227). Consider two sets of particles, one `merely' shaped as
a hammer, the other shaped as a hammer {\it and} bonded together so as
to maintain its shape under the stress of, for example, hitting a
nail.  The latter is a hammer, the former not.  Bedard then points out
that the pilot-wave theory may account for such `bonds' by appeal to
features of the wavefunction, but {\it not} by appeal to the intrinsic
properties of the particles, of which, on the minimalist account,
there is just one---position.  Hence the minimalist interpretation is
inadequate, for it {\it must} model the distinction between bonded and
unbonded particles and yet cannot: ``quantum mechanics would not be
celebrated for successfully modeling such a distinction if the
distinction were insignificant'' (p.~228).  (We shall see that the
minimalist is not much inclined to join the celebration.)

Bedard's second argument is aimed at the idea that the configuration
of particles could `cause' our perceptions.  In partiular, she
considers a quantum-mechanical evolution such as:
\begin{equation}
\label{eq:ev1}
  \ket{v}\ket{\mbox{``}v\mbox{''}}\ket{\mbox{ready}} 
  \rightarrow
  \ket{v}\ket{\mbox{``}v\mbox{''}}\ket{\mbox{sees ``$v$''}}
\end{equation}
where $\ket{v}$ is the state of some measured system,
$\ket{\mbox{``}v\mbox{''}}$ is the state of an apparatus indicating
the result ``$v$'', $\ket{\mbox{ready}}$ is the state of an observer
not yet having observed the apparatus, and $\ket{\mbox{sees ``$v$''}}$
is the state of an observer having observed the apparatus.  Bedard
then notes that the evolution of the particles in the observer are not
functionally dependent on the positions of the apparatus' particles.

We are supposed to conclude that the apparatus' particles do not,
therefore, `cause' the perception ``sees `v'\,''.  To bolster this
conclusion, Bedard further claims that on the counterfactual analysis
of causation, the positions of the apparatus' particles do not affect
the observer's particles, because ``according to the counterfactual
analysis, `the pointer particles cause the perception' means that if
the pointer configuration were different, then the perception would
have been different'' (p.~231).  But we should be somewhat more
careful, here.  Surely, for example, the counterfactual analysis says
more.  Consider a baseball breaking a window.  Is it true that `had
the ball been elsewhere, the window would not have broken'?  No.  One
requires that the difference be `enough'.  (E.g., $10^{-10}$cm is
probably not enough.)  True, the configurations that {\it might} make
a difference happen to have probability zero, if in fact the
wavefunction is as Bedard describes in (\ref{eq:ev1}), but several
problems arise here.  First, is probability $0$ low enough?  If not,
how do we evaluate the relevant counterfactual?  If so, are we allowed
to alter the wavefunction (in order to make these configurations to
have probability greater than $0$)?  Such questions make it clear that
appeal to the counterfactual analysis is at best problematic, and
requires considerably more careful discussion.  (See
\citep{dickson96a} for some discussion of the difficulties for
applying the counterfactual analysis to the pilot-wave theory.)

In fact, Bedard does briefly raise the possibility that
the evolution is, instead,
\begin{equation}
\label{eq:ev2}
  \Big(\ket{v}\ket{\mbox{``}v\mbox{''}} + \ket{w}\ket{\mbox{``}w\mbox{''}}\Big)\ket{\mbox{ready}} 
  \rightarrow
  \ket{v}\ket{\mbox{``}v\mbox{''}}\ket{\mbox{sees ``$v$''}} 
    + \ket{w}\ket{\mbox{``}w\mbox{''}}\ket{\mbox{sees ``$w$''}}.
\end{equation}
In this case, one or the other of the wavepackets for the apparatus is
`empty' (i.e., the configuration is not located there)---let it be
$\ket{\mbox{``}w\mbox{''}}$.  Then we might be tempted to say that if
the configuration of the apparatus {\it had} been there, the observer
{\it would have} seen ``$w$'' rather than ``$v$''.  But then does the
counterfactual analysis not entail that the configuration is causally
relevant to the state of the observer?  Bedard answers `no', for two
reasons.  First, ``the viability of particularity [i.e., the minimal
interpretation] \ldots should not hinge on the complexity of the
universe'' (p.~232) and second, ``we could construct more realistic
and complex examples in which empty wavepackets exist without having
the pointer particles determine which wavepacket is active or affect
the brain particles' trajectories'' (ibid.).  The first point is
crucial---I shall suggest below that in fact the only reasonable
version of the minimal interpretation {\it should} allow that answers
to `causal' questions depend on the contingent details of this
universe.  The second, I do not understand, if Bedard has in mind the
situation that I described schematically in (\ref{eq:ev2}).  There,
the quantum-mechanical perfect correlations between the apparatus and
the observer guarantee that the observer will (with probability $1$)
see the result that the apparatus indicates.  However, this latter
point is central neither to her argument nor to mine.  I shall address
the general argument about `particulate epistemology' below.

Bedard's third argument is that mere positions are insufficient to
explain the correlation between our perceptions and the world.  For
example, colors may not depend on configuration in the appropriate
way, and yet we can perceive color.  In general, it is possible to
perceive properties that are not in any obvious way dependent on
configurations.  But then configurations cannot explain our
perceptions.

\section{The Minimal Interpretation}

Bedard quotes several authors who seem to adopt something like a
`minimalist' interpretation of the pilot-wave theory.  We will do
well, nonetheless, to make it clear what that interpretation says.

Bohm's formulation of the pilot-wave theory in 1952 invoked a `quantum
potential' in addition to the classical potential, and toegether they
are responsible for the motions of particles---they are the potentials
that appear in Hamilton's equations.  Bohm seems never to
have let go of the idea of the quantum potential \citep{bohm93a}.  But
why is it necessary?  Apparently it is required to explain the
`deviation' of particles from Newtonian trajectories.  For example, in
the two-slit experiment, where there is no classical force acting on
the particle between the slits and the screen, one `must' invoke a
`quantum force' to explain why the particle does not follow the
Newtonian trajectory.  (\citet{bohm93a} provide some nice pictures of
these `curved' trajectories.)

The minimalist interpretation that I would advocate (were I an
advocate of the pilot-wave theory in the first place) begins by asking
why we must appeal to Newton to establish `what is expected'.
Instead, why not simply continue to allow that the particle
experiences no force between the slits, and yet its trajectory is not
the classically expected trajectory?  This idea suggests that we
consider a space-time in which these non-classical yet free motions
are geodesics, so that, in much the same way that we no longer invoke
the `force of gravity' to explain deviations from Euclidean geodesics,
so also we would not invoke the `quantum potential' to explain
deviations from the Newtonian trajectories.  This idea has been
carried out with enough rigor to render it at least a plausible
foundation for an interpretation of the pilot-wave theory
\citep{pitowsky91a}.

The sole role for the wavefunction, then, is to determine the
structure of space-time.  It does not describe any other `real
features' of the world, and there are no `forces', `potentials', or
anything of the sort accounting for the non-classicality of the
motions of particles.

What sort of metaphysics goes with this view?  It is reductionistic,
in the traditional sense.  A crucial part of this attitude of
reductionism is that our accounts of the world---in particular, the
categories that we use to describe the world and relations amongst
objects in it---might not be `fundamental', but might instead be
imposed by us on the world.  A familiar example is afforded by heat.
Suppose that statistical mechanics is a successful reduction of
thermodynamics, that heat is nothing more than the motion of
molecules, and similarly for the other central concepts of
thermodynamics.  Then, where thermodynamics might lead us to refer to
such things as `the quantity of heat', `the flow of heat', and so
forth, we would, in light of this reduction, understand that these
phrases do not straightforwardly refer to any real entities or
properties of entities in the world.

An example familiar to philosophers is afforded by Hume's account of
causation.  Causes are not in the world, according to this account;
they are `constant conjunction with an inference by the mind'; i.e.,
we impose causal relations on the world, where in fact there is
nothing other than constant conjunction.  In general, the minimalist
interpretation is open to the idea that the categories that you and I use
to describe the world are not the real categories into which the
things in the world actually fall---not even close.  The minimalist is
open to a radical revision of our ordinary discourse about the world
in light of our interpretations of scientific theory.

\section{Replies to Bedard's Arguments}

This feature of the minimal interpretation---and reductionism in
general---makes it clear where Bedard's arguments miss the mark.  She
assumes the legitimacy of certain distinctions made {\it by us}, and
then requires that physical theory provide an explanation of, or
account of, those distinctions in purely physical terms.  The
minimalist, however, is open to the idea that our mode of description
may be largely responsible for the putative distinctions.  While
Bedard does implicitly acknowledge that her arguments do not apply to
this reductionistic form of minimalism (I shall quote two cases
below), apparently she does not sufficiently appreciate that
reductionism was the only plausible form of minimalism in the first
place.

Consider again the case of the real and false hammers.  Bedard notes
that there are no resources internal to the minimalist's description
of the world to distinguish between real and false hammers.  But the
minimalist need not agree that such distinctions are reflected in
fundamental physical facts about the world.  As far as the world is
concerned, `merely' hammer-shaped sets of particles and `true hammers'
are no different---in much the same way that there is no physically
fundamental difference between a beautiful painting and an ugly one.
The sets of particles that {\it happen}, by virtue of their
trajectories, to remain shaped as a hammer even under `stress'
(another concept imposed by us!) are picked out {\it by us} as
special---presumably because they are useful for pounding nails.

It is crucial, now, to note that the pilot-wave theory {\it does}
predict that, under the right conditions (that is, in a universe with
the right sort of spatio-temporal structure and initial
configuration---and we may have just gotten `lucky' in this respect),
there will be sets of particles that are hammer-shaped, and that will
remain so under `stress'.  It is crucial, in other words, to realize
that Bedard has not shown that there could be no hammers, according to
the minimalist view.  Her argument, rather, establishes that the
minimalist view has no {\it explanation} of their `hammerhood',
rather than their `hammer-shapedness'.  In general, it has no {\it
explanation}, in terms of fundamental physical facts, of the
difference between `bonded' and `unbonded' particles.  But the
minimalist interpretation need provide no such explanation.

The virtue of Bedard's first argument is to highlight this fact in a
particularly sharp way, and to make it clear that the minimalist must
be a radical reductionist of roughly the Humean sort.  But the
minimalist was already committed to this view from the start---one can
hardly claim that the only truly existent objects are point particles
with positions and fail to notice that such a claim involves a
particularly radical form of reductionism.  Some may be unsatisfied by
the view, but Bedard has not shown that it ``do[es] not make sense''
(p.~223).  To put it in her terms, Bedard {\it has} demonstrated ``the
incompatibility of particularity with theories in which certain
material objects have essential properties that are causal'' (p.~229),
but the minimalist need acknowledge no such objects.

What I have said to this point should make it clear how to respond to
Bedard's second argument as well.  The point there (and let us grant it
in spite of the problematic invocation of the counterfactual analysis
of causation) was that configurations do not, in general, cause
observers' perceptions.  Again, the reductionist is not committed to
the view that configurations {\it do} cause observers' perceptions.
Indeed, the very notion of a causal conection is dispensible on this
view.  The minimalist interpretation {\it does} predict the requisite
correlations.  Non-reductionists understand them causally.

Bedard's third argument contains the germ of a potential
problem for the minimalist.  However, outside the context of some
concrete theory of perception---not to mention our consciousness of
perceptions---there is very little that can (or ought to be) said
about the relation between the physical world and our perceptions of
it.  Nonetheless, the potential problem is that if this theory, `we
know not what', entails that consciousness has nothing to do with
configurations, then the minimalist will have some explaining to do.
One might be tempted to add that Bedard has shown that {\it however}
our brains physically encode information about the world, the
configurations of particles in the world cannot be the causes of those
encodings.  Agreed, but we have already seen that the minimalist need
not provide a description of such putative causal connections in
physically fundamental terms (i.e., purely in terms of the positions
of point particles).

\section{Objection and Conclusion}

One might object that the position I am describing here is just the
position already examined by Bedard, namely, that the hammers are
distinguished from the non-hammers not by their instantaneous
properties, but by their entire histories (trajectories).  She rejects
this view, saying that ``if an object's essential properties include
causal properties, these causal properties should not be smuggled in
through some of their effects (such as particle trajectories)''
(p.~229).  The reductionists as I have described them will, of course,
reject the antecedent.

There are two ways to do so.  First, the reductionist might indeed be
able to reduce causal properties to properties of a trajectory, in
which case causal properties are not in fact `essential', in the same
way that heat is not an essential property of ensembles of particles
(let us assume).  Having reduced heat to the motion of molecules, it
is no good saying ``if heat is an essential property, then do not
smuggle it in through its effects (the motions of molecules)'' for
heat is not essential.  Second, the reductionist might simply refuse
to admit that causal properties are even reducible to fundamental
physical properties.  Causal properties simply have no place at all in
a physical theory.  So said Hume: the property of being a hammer has
two parts, `constant hammer-shape' and `inference by the mind'.

So what do we learn from Bedard's paper?  We learn that the minimalist
{\it must} be a reductionist.  That lesson is valuable, though I have
suggested that in any case there was little doubt, even prior to
Bedard's paper, that the minimalist should be a reductionist.
Nonetheless, Bedard has provided us a very fine illustration, in detail,
of just why the minimalist must be a reductionist, and just how
radical that reductionism might have to be.

\bibliography{bohm.bib}

\begin{thebibliography}{}

\bibitem[Bedard, 1999]{bedard99a}
Bedard, K. (1999).
\newblock {M}aterial {O}bjects in {B}ohm's {I}nterpretation.
\newblock {\em Philosophy of Science}, 66:221--242.

\bibitem[Bohm, 1952]{bohm52a}
Bohm, D. (1952).
\newblock {A} {S}uggested {I}nterpretation of the {Q}uantum {T}heory in {T}erms
  of `{H}idden {V}ariables'.
\newblock {\em Physical Review}, 85:166--193.

\bibitem[Bohm and Hiley, 1993]{bohm93a}
Bohm, D. and Hiley, B. (1993).
\newblock {\em {T}he Undivided Universe: An Ontological Interpretation of
  Quantum Theory}.
\newblock Routledge, New York.

\bibitem[de~Broglie, 1924]{debroglie24a}
de~Broglie, L. (1924).
\newblock {\em {R}echerches sur la theories des quanta}.
\newblock PhD thesis, Universit\'{e} de Paris.

\bibitem[Dickson, 1996]{dickson96a}
Dickson, M. (1996).
\newblock {I}s the {B}ohm {T}heory {L}ocal?
\newblock In Cushing, J., Fine, A., and Goldstein, S., editors, {\em Bohmian
  Mechanics: An Appraisal}. Kluwer, Dordrecht.
\newblock Vol.~184 of {\em Boston Studies in Philosophy of Science}.

\bibitem[Pitowsky, 1991]{pitowsky91a}
Pitowsky, I. (1991).
\newblock {B}ohm {Q}uantum {P}otentials and {Q}uantum-{G}ravity.
\newblock {\em Foundations of Physics}, 21:343--352.

\end{thebibliography}

\end{document}